\documentstyle[aps,prd,amssymb]{revtex}

\newcommand{\be}{\begin{equation}}
\newcommand{\ee}{\end{equation}}
\newcommand{\bea}{\begin{eqnarray}}
\newcommand{\eea}{\end{eqnarray}}

\def\npi{({\bf n}_{12}\cdot{\bf p}_1)}
\def\pipi{{\bf p}_1^2}
\def\pipip{({\bf p}_1^2)}

\def\npj{({\bf n}_{12}\cdot{\bf p}_2)}
\def\pjpj{{\bf p}_2^2}
\def\pjpjp{({\bf p}_2^2)}

\def\pipj{({\bf p}_1\cdot{\bf p}_2)}

\def\omk{\omega_{\text{kinetic}}}
\def\oms{\omega_{\text{static}}}

\begin{document}

\title{Poincar\'e invariance in the ADM Hamiltonian approach
to the general relativistic two-body problem}

\author{Thibault Damour}
\address{Institut des Hautes \'Etudes Scientifiques,
91440 Bures-sur-Yvette, France}

\author{Piotr Jaranowski}
\address{Institute of Theoretical Physics,
University of Bia{\l}ystok,
Lipowa 41, 15-424 Bia{\l}ystok, Poland}

\author{Gerhard Sch\"afer}
\address{Theoretisch-Physikalisches Institut,
Friedrich-Schiller-Universit\"at,
Max-Wien-Pl.\ 1, 07743 Jena, Germany}

\maketitle

\begin{abstract}
A previously found momentum-dependent regularization ambiguity in the third
post-Newtonian two point-mass Arnowitt-Deser-Misner Hamiltonian is shown to be
uniquely determined by requiring global Poincar\'e invariance. The phase-space
generators realizing the Poincar\'e algebra are explicitly constructed.

\vspace{2ex}\noindent
PACS number(s): 04.25.Nx, 04.20.Fy, 04.30.Db, 97.60.Jd
\end{abstract}

\vspace{2ex}

The equations of motion of a gravitationally interacting two point-mass system
have been derived some years ago up to the 5/2 post-Newtonian (2.5PN)
approximation\footnote{We recall that the ``$n$PN approximation'' refers to the
terms of order $(v/c)^{2n}\sim\left(GM/(c^2r)\right)^n$ in the equations of
motion.}, in harmonic coordinates \cite{DD,D82,LH}.  Recently, it has been
possible to derive the third post-Newtonian (3PN) Hamiltonian of a two
point-mass system \cite{JS98} within the canonical formalism of Arnowitt, Deser
and Misner (ADM) \cite{ADM}.  It was found that, at the 3PN level, the use of
Dirac-delta-function sources to model the two-body system causes the appearance
of badly divergent integrals which (contrary to what happened at the 2.5PN
\cite{LH,S85} and 3.5PN \cite{JS97} levels) cannot be unambiguously regularized
\cite{JS98,JS99,DJS1}.  The ambiguities in the regularization of the 3PN
divergent integrals are parametrized by two quantities:  $\oms$ and $\omk$.

Prompted by a recent remark \cite{JSWeimar}, the purpose of this work is to show
that requiring the (global) Poincar\'e invariance of the 3PN ADM Hamiltonian
dynamics uniquely determines one (and only one) of these regularization
ambiguities:  namely the ``kinetic ambiguity'' parameter $\omk$.  [The ``static
ambiguity'' $\oms$ remains unconstrained because it parametrizes a
${\cal O}(c^{-6})$ Galileo-invariant additional contribution to the 3PN 
Hamiltonian.] Parallel work in the harmonic-coordinates approach to 3PN dynamics 
has recently obtained similar results \cite{BF}.

Note that general relativity admits (when considering isolated systems) the full 
Poincar\'e group as a {\it global} symmetry. Therefore, whatever be the 
coordinate system used (as long as it respects asymptotic flatness) the 
general relativistic dynamics of $N$-body systems should embody some 
representation of this global Poincar\'e symmetry. When solving Einstein's 
equation by a weak-field, ``post-Minkowskian'' expansion,
$\sqrt{g}\,g^{\mu \nu}-\eta^{\mu\nu} \equiv h^{\mu \nu}
= G \, h_{(1)}^{\mu \nu} + G^2 \, h_{(2)}^{\mu \nu} + \cdots$,
and fixing the gauge by the ``harmonicity 
condition'', $\partial_{\nu} \, h^{\mu \nu} = 0$, the whole scheme stays 
manifestly invariant under the usual (linear) representation of the 
Poincar\'e group: $x'^{\mu} = \Lambda_{\nu}^{\mu} \, x^{\nu} + a^{\mu}$ 
(assuming that the regularization procedure used to deal with the point-mass 
divergencies is manifestly Poincar\'e invariant). In such a case the 
$N$-body dynamics will be invariant under the representation of the 
Poincar\'e group induced on the dynamical variables, say $x_a^i (t)$, 
$\dot{x}_a^i (t)$, $a=1,\ldots ,N$, by the action of the usual linear 
Poincar\'e transformations. This global Poincar\'e symmetry has been 
explicitly checked at the 2PN level in Ref.\ \cite{DD81b} by proving that 
the 2PN (acceleration-dependent) two-body Lagrangian in harmonic 
coordinates \cite{DD81a,D82} changed only by a total time derivative under a 
generic, infinitesimal Poincar\'e transformation. In this work we consider 
the 3PN two-body Hamiltonian derived by Ref.\ \cite{JS98} within the ADM 
canonical formalism. This formalism is not manifestly Poincar\'e invariant 
because it splits space and time, and fixes the coordinates by the following 
gauge conditions: $\delta_{ij} \, \pi^{ij} = 0$, $\partial_j \left( g_{ij} - 
\frac{1}{3} \, g_{ss} \, \delta_{ij} \right) =  0$. This lack of {\it 
manifest} Poincar\'e invariance is not problematic (though it introduces 
some technical complications). Indeed, we shall explicitly show in this 
paper that the global Poincar\'e symmetry of the two-body dynamics can be 
realized in phase space, albeit by a somewhat complicated, nonlinear action.

The basic principle that we shall follow to study Poincar\'e invariance of 
the 3PN two-body Hamiltonian $H ({\bf x}_a,{\bf p}_a)$, $a=1,2$,
with its associated Poisson bracket structure,
\begin{equation} 
\label{eq1}
\{A({\bf x}_a,{\bf p}_a),B ({\bf x}_a,{\bf p}_a)\} \equiv \sum_a \sum_i 
\left( \frac{\partial A}{\partial \, x_a^i} \, \frac{\partial B}{\partial \, 
p_{ai}} - \frac{\partial A}{\partial \, p_{ai}} \, \frac{\partial B}{\partial 
\, x_a^i} \right) \, , 
\end{equation}
is the following: the presence of a Poincar\'e symmetry is equivalent to 
requiring the existence of ``generators'' $P^{\mu}$, $J^{\mu \nu}$, realized 
as functions $P^{\mu}({\bf x}_a,{\bf p}_a)$, $J^{\mu \nu}({\bf x}_a,{\bf p}_a)$ 
on the 
two-body phase-space $({\bf x}_1,{\bf x}_2,{\bf p}_1,{\bf p}_2)$, whose Poisson 
brackets 
(\ref{eq1}) satisfy the usual Poincar\'e algebra (here we set $c=1$):
\begin{mathletters}
\label{eq2}
\begin{eqnarray}
\{ P^{\mu} , P^{\nu} \} &= &0 \, , \label{eq2a} \\[2ex]
\{ P^{\mu} , J^{\rho \sigma} \} &= &-\eta^{\mu \rho} \, P^{\sigma} + 
\eta^{\mu \sigma} \, P^{\rho} \, , \label{eq2b} \\[2ex]
\{ J^{\mu \nu} , J^{\rho \sigma} \} &= &-\eta^{\nu \rho} \, J^{\mu \sigma} + 
\eta^{\mu \rho} \, J^{\nu \sigma} + \eta^{\sigma \mu} \, J^{\rho \nu} - 
\eta^{\sigma \nu} \, J^{\rho \mu} \label{eq2c},
\end{eqnarray}
\end{mathletters}
where $\eta_{\mu \nu} = {\rm diag}\,(-1,+1,+1,+1)$.

The functions $P^{\mu} ({\bf x}_a , {\bf p}_a)$, 
$J^{\mu \nu} ({\bf x}_a , {\bf p}_a)$ generate (in 
phase space) the infinitesimal Poincar\'e transformations: $\delta_{\alpha , 
\omega} \, F = \{ F , \alpha^{\mu} \, P_{\mu} + \frac{1}{2} \, \omega^{\mu 
\nu} \, J_{\mu \nu} \}$. Finite transformations are then (in principle) 
defined by exponentiating these infinitesimal actions. The satisfaction of 
the algebra (\ref{eq2}) ensures that one thereby generates a consistent 
Poincar\'e symmetry. The time component $P^0$ (i.e. the total energy) is 
realized as the Hamiltonian $H ({\bf x}_a , 
{\bf p}_a)$ (including the rest-mass contribution). The other 
generators can be decomposed as: $P^i$ (three momentum), $J^i \equiv 
\frac{1}{2} \, \varepsilon^{ik\ell} \, J_{k\ell}$ (angular momentum), and 
$K^i \equiv J^{i0}$ (boost vector). One further decomposes the boost vector 
$K^i$ 
(which represents the constant of motion associated to the center of mass 
theorem) as
$K^i ({\bf x}_a , {\bf p}_a ; t) \equiv G^i ({\bf x}_a , {\bf p}_a) - t \, P^i 
({\bf x}_a , {\bf p}_a)$
so that the total time derivative $dK^i / dt = \partial K^i / \partial t + \{ 
K^i , H \} = -P^i + 
\{ G^i , H \} = 0$. Finally, the Poincar\'e algebra explicitly reads
\begin{mathletters}
\bea
\{ P_i , H \} = \{ J_i , H \} = 0 \, , \label{eq3}
\\[2ex]
\{ J_i , P_j \} = \varepsilon_{ijk} \, P_k \, , \ \{ J_i , J_j \} = 
\varepsilon_{ijk} \, J_k \, , \label{eq4}
\\[2ex]
\{ J_i , G_j \} = \varepsilon_{ijk} \, G_k \, , \label{eq5}
\\[2ex]
\{ G_i , H \} = P_i \, , \label{eq6}
\\[2ex]
\{ G_i , P_j \} = \frac{1}{c^2} \, H \, \delta_{ij} \, , \label{eq7}
\\[2ex]
\label{eq8}
\{ G_i , G_j \} = - \frac{1}{c^2} \, \varepsilon_{ijk} \, J_k \, . 
\eea
\end{mathletters}

As the gauge fixing used in the ADM formalism manifestly respects the Euclidean 
group (which means that $H ({\bf x}_a,{\bf p}_a)$ is translationally and 
rotationally invariant), the generators $P_i$ and $J_i$ are simply realized as
\begin{mathletters}
\bea
\label{eq9}
P_i ({\bf x}_a , {\bf p}_a) &=& \sum_a p_{ai} \, , 
\\[2ex]
\label{eq10}
J_i ({\bf x}_a , {\bf p}_a) &=& \sum_a
\varepsilon_{ik\ell} \, x_a^k \, p_{a\ell},
\eea
\end{mathletters}
and exactly satisfy Eqs.\ (\ref{eq3}) and (\ref{eq4}). The condition 
(\ref{eq5}) will also be exactly satisfied if $G_i$ is constructed as a 
three-vector from ${\bf x}_a$ and ${\bf p}_a$. 
Finally, the condition for full Poincar\'e invariance boils down to the 
existence of a vector $G_i({\bf x}_a,{\bf p}_a)$ 
satisfying the three non-trivial relations (\ref{eq6}), (\ref{eq7}) and 
(\ref{eq8}), in which enters, besides $P_i$ and $J_i$ given in 
Eqs.\ (\ref{eq9}) and (\ref{eq10}), the full (3PN-accurate) Hamiltonian:
\be
\label{eq11}
H ({\bf x}_a , {\bf p}_a) = \sum_a m_a c^2
+  H_{\text{N}}({\bf x}_a , {\bf p}_a)
+ \frac{1}{c^2} \, H_{\rm 1PN} ({\bf x}_a , {\bf p}_a)
+ \frac{1}{c^4} \, H_{\rm 2PN} ({\bf x}_a , {\bf p}_a) 
+ \frac{1}{c^6} \, H_{\rm 3PN}({\bf x}_a , 
{\bf p}_a) + {\cal O} \left( \frac{1}{c^8} \right) \, . 
\ee
At the Newtonian order, i.e.\ when keeping the rest-mass term 
$\Sigma_a m_a c^2$ and the Newtonian-level Hamiltonian,
\be
\label{eq12}
H_{\text{N}}({\bf x}_a , {\bf p}_a) = \sum_a \, 
\frac{{\bf p}_a^2}{2\,m_a} - \frac{1}{2} \sum_a \sum_{b \ne a} 
\frac{G \, m_a \, m_b}{r_{ab}} \, , 
\ee
($r_{ab} \equiv \vert {\bf x}_a - {\bf x}_b \vert$), 
it is easily checked that the usual Newtonian center-of-mass vector
\begin{equation}
G_N^i ({\bf x}_a , {\bf p}_a) \equiv \sum_a m_a \, 
x_a^i \label{eq13}
\end{equation}
satisfies Eqs.\ (\ref{eq6})--(\ref{eq8}). [Note that, in this approximation, 
the right-hand side of Eq.\ (\ref{eq7}) yields
$\left(\sum_{a}m_a\right)\delta_{ij}$
from the rest-mass contribution to $H$.]

To study the existence of $G^i$ beyond the Newtonian approximation, we need 
the explicit expressions of the 1PN, 2PN and 3PN contributions to 
the Hamiltonian (\ref{eq11}) in an arbitrary reference frame. The 1PN 
contribution,
\bea
\label{eq14}
H_{\text{1PN}}({\bf x}_a,{\bf p}_a) &=&
- \frac{1}{8}\frac{\pipip^2}{m_1^3}
+ \frac{1}{8}\frac{Gm_1m_2}{r_{12}} \left[
- 12\,\frac{\pipi}{m_1^2}
+ 14\,\frac{\pipj}{m_1m_2}
+ 2\,\frac{\npi\npj}{m_1m_2} \right]
\nonumber\\[2ex]&&
+ \frac{1}{4}\frac{Gm_1m_2}{r_{12}}\frac{G(m_1+m_2)}{r_{12}}
+ \biglb(1\longleftrightarrow 2\bigrb),
\eea
has been known for a long time.
The operation ``$+\biglb(1\longleftrightarrow 2\bigrb)$'' in Eq.\ (\ref{eq14})
denotes the addition for each term in Eq.\ (\ref{eq14}) (including the ones
which are symmetric under label exchange) of another term obtained
by the label permutation $1\longleftrightarrow 2$.
The 2PN-accurate explicit expression of 
$H({\bf x}_a ,{\bf p}_a)$, in the ADM formalism, 
was derived in Ref.~\cite{DS88} (Eq.~(2.5) there). [The corresponding 
explicit Lagrangian $L_{2 \, {\rm PN}}^{\rm ADM} ({\bf x}_a , 
\dot{\bf x}_a)$ is given in Ref.~\cite{DS85}.] These results 
corrected earlier results by Ohta et al.\ \cite{OOKH73}. The final result 
reads:
\bea
\label{eq15}
H_{\text{2PN}}({\bf x}_a,{\bf p}_a) &=&
\frac{1}{16}\frac{\pipip^3}{m_1^5}
+ \frac{1}{8} \frac{Gm_1m_2}{r_{12}} \left[
5\,\frac{\pipip^2}{m_1^4}
- \frac{11}{2}\frac{\pipi\,\pjpj}{m_1^2m_2^2}
- \frac{\pipj^2}{m_1^2m_2^2}
+ 5\,\frac{\pipi\,\npj^2}{m_1^2m_2^2}
\right.\nonumber\\[2ex]&&\left.
- 6\,\frac{\pipj\,\npi\npj}{m_1^2m_2^2}
- \frac{3}{2}\frac{\npi^2\npj^2}{m_1^2m_2^2} 
\right]
\nonumber\\[2ex]&&
+ \frac{1}{4}\frac{G^2m_1m_2}{r_{12}^2} \left[ 
m_2\left(10\frac{\pipi}{m_1^2}+19\frac{\pjpj}{m_2^2}\right)
- \frac{1}{2}(m_1+m_2)\frac{27\,\pipj+6\,\npi\npj}{m_1m_2} \right]
\nonumber\\[2ex]&&
- \frac{1}{8} \frac{Gm_1m_2}{r_{12}}\frac{G^2 (m_1^2+5m_1m_2+m_2^2)}{r_{12}^2}
+ \biglb(1\longleftrightarrow 2\bigrb).
\eea

Ref.\ \cite{JS98} derived the 3PN-accurate ADM Hamiltonian {\it restricted to
the center-of-mass reference frame}:  ${\bf p}_1+ {\bf p}_2=0$.  For the present
work we have generalized Ref.\ \cite{JS98} in deriving $H_{\rm 3PN}$ in an
arbitrary reference frame.  Our starting point for doing this calculation is the
improved form of the 3PN Hamiltonian, $\widetilde{H}_{\rm 3PN}$, given in
Appendix A of Ref.\ \cite{DJS1} (Eqs.\ (A8)--(A10) there).  Note first that
$\widetilde{H}_{3 \, {\rm PN}}$ defined there denotes the {\it higher-order}
Hamiltonian $\widetilde{H}_{\rm 3PN}({\bf x}_a ,{\bf p}_a,\dot{\bf x}_a,\dot{\bf 
p}_a)$
defined by eliminating the field variables $h_{ij}^{\rm TT}$,
$\dot{h}_{ij}^{\rm TT}$ in the ``Routh functional''
$R({\bf x}_a ,{\bf p}_a ,h_{ij}^{\rm TT},\dot{h}_{ij}^{\rm TT})$
introduced in Eq.\ (33) of Ref.\ \cite{JS98}.  However, it was shown in
Ref.\ \cite{DJS1} that one could {\it reduce} the higher-order Hamiltonian
$\widetilde{H}_{\rm 3PN}({\bf x}_a ,{\bf p}_a, \dot{\bf x}_a , \dot{\bf p}_a)$
to an {\it ordinary} Hamiltonian $H_{\rm 3PN}({\bf x}'_a,{\bf p}'_a)$,
at the price of the following (3PN-level) shift of phase-space coordinates:
\be
{\bf x}'_a = {\bf x}_a + \frac{\partial \widetilde 
H}{\partial \, \dot{\bf p}_a} \quad , \quad {\bf p}'_a 
= {\bf p}_a - \frac{\partial \widetilde H}{\partial \, 
\dot{\bf x}_a} \, . \label{eq16}
\ee
After performing the shift (\ref{eq16}) with respect to the original ADM 
coordinates ${\bf x}_a$, ${\bf p}_a$ (we henceforth 
drop the primes for notational simplicity), the calculation of the 3PN 
(order-reduced) Hamiltonian consists in evaluating three very complicated 
integrals: 
\begin{equation}
H_{\rm 3PN} = - \frac{5}{128} \, \sum_a \, 
({\bf p}_a^2)^4 + \int d^3 x (h_1^{\rm red} + h_2 + h_3) \, . 
\label{eq16b}
\end{equation}
The integrands $h_1$, $h_2$, $h_3$ are given in Eqs.\ (A9) of \cite{DJS1}. The 
order-reduced integrand $h_1^{\rm red}$ is defined (as shown in \cite{DJS1}) 
by using the Newtonian equations of motion to eliminate 
$\dot{\bf x}_a$ and $\dot{\bf p}_a$ when 
computing the time derivative $\dot{h}_{ij}^{\rm TT}$ (which enters the last 
two terms of $h_1$). As explained in \cite{DJS1}, this new form of the 3PN 
Hamiltonian is free of ``contact term'' ambiguities, and the integrals it 
contains can all be uniquely defined by using the {\it Riesz-type 
regularization} procedure explained in \cite{JS98}. We have recomputed from 
scratch all the integrals by using the generalized Riesz formula given in 
\cite{JS98}. This {\it Riesz-regularized} 3PN Hamiltonian reads explicitly 
(in an arbitrary reference frame)\footnote{The authors thank Luc Blanchet for 
pointing out that two terms were missing in the printed version of our 
non-center-of-mass Hamiltonian.}
\bea
\label{eq17}
H_{\text{3PN}}^{\rm reg}({\bf x}_a,{\bf p}_a) &&=
-\frac{5}{128}\frac{\pipip^4}{m_1^7}
+ \frac{1}{32} \frac{Gm_1m_2}{r_{12}} \Bigg[
- 14\,\frac{\pipip^3}{m_1^6}
+ 4\,\frac{\biglb(\pipj^2+4\,\pipi\,\pjpj\bigrb)\pipi}{m_1^4m_2^2}
+ \frac{\biglb(\pipi\,\pjpj-2\,\pipj^2\bigrb)\pipj}{m_1^3m_2^3}
\nonumber\\[2ex]&&
- 10\,\frac{\biglb(\pipi\,\npj^2+\pjpj\,\npi^2\bigrb)\pipi}{m_1^4m_2^2}
+ 24\,\frac{\pipi\,\pipj\npi\npj}{m_1^4m_2^2}
+ 2\,\frac{\pipi\,\pipj\npj^2}{m_1^3m_2^3}
\nonumber\\[2ex]&&
+ \frac{\biglb(7\,\pipi\,\pjpj-10\,\pipj^2\bigrb)\npi\npj}{m_1^3m_2^3}
+ 6\,\frac{\pipi\,\npi^2\npj^2}{m_1^4m_2^2}
\nonumber\\[2ex]&&
+ 15\,\frac{\pipj\npi^2\npj^2}{m_1^3m_2^3}
- 18\,\frac{\pipi\,\npi\npj^3}{m_1^3m_2^3}
+ 5\,\frac{\npi^3\npj^3}{m_1^3m_2^3} \Bigg]
\nonumber\\[2ex]&&
+ \frac{G^2m_1m_2}{r_{12}^2} \Bigg[
\frac{1}{16}(m_1-27m_2)\frac{\pipip^2}{m_1^4}
- \frac{115}{16}m_1\frac{\pipi\,\pipj}{m_1^3m_2}
+ \frac{1}{48}m_2\frac{25\,\pipj^2+371\,\pipi\,\pjpj}{m_1^2 m_2^2}
\nonumber\\[2ex]&&
+ \frac{17}{16}\frac{\pipi\npi^2}{m_1^3}
- \frac{1}{8}m_1 
\frac{\biglb(15\,\pipi\,\npj+11\,\pipj\,\npi\bigrb)\npi}{m_1^3 m_2}
+ \frac{5}{12}\frac{\npi^4}{m_1^3}
\nonumber\\[2ex]&&
- \frac{3}{2}m_1\frac{\npi^3\npj}{m_1^3m_2}
+ \frac{125}{12}m_2\frac{\pipj\,\npi\npj}{m_1^2m_2^2}
+ \frac{10}{3}m_2\frac{\npi^2\npj^2}{m_1^2m_2^2}
\nonumber\\[2ex]&&
- \frac{1}{48} (220 m_1 + 193 m_2) \frac{\pipi \npj^2}{m_1^2 m_2^2}
\Bigg]
+ \frac{G^3m_1m_2}{r_{12}^3} \Bigg[
-\frac{1}{48}
\bigglb(466\,m_1^2+\Big(473-\frac{3}{4}\pi^2\Big)m_1m_2+150\,m_2^2\biggrb)
\frac{\pipi}{m_1^2}
\nonumber\\[2ex]&&
+ \frac{1}{16}
\bigglb(77(m_1^2+m_2^2)+\Big(143-\frac{1}{4}\pi^2\Big)m_1m_2\biggrb)
\frac{\pipj}{m_1m_2}
+ \frac{1}{16}
\bigglb(61\,m_1^2-\Big(43+\frac{3}{4}\pi^2\Big)m_1m_2\biggrb)
\frac{\npi^2}{m_1^2}
\nonumber\\[2ex]&&
+ \frac{1}{16}
\bigglb(21(m_1^2+m_2^2)+\Big(119+\frac{3}{4}\pi^2\Big)m_1m_2\biggrb)
\frac{\npi\npj}{m_1m_2} \Bigg]
\nonumber\\[2ex]&&
+ \frac{1}{8} \frac{G^4m_1m_2^3}{r_{12}^4}
\Bigg[ \bigg(\frac{227}{3}-\frac{21}{4}\pi^2\bigg)m_1+m_2 \Bigg]
+ \biglb(1\longleftrightarrow 2\bigrb).
\eea
However, it was emphasized in Refs.\ \cite{JS98,JS99,DJS1} 
that the nature of the divergent integrals which had to be regularized to 
compute $H_{\rm 3PN}^{\rm reg}$, Eq.\ (\ref{eq17}), was such that the 
result should be considered as being partly ambiguous. These {\it 
regularization ambiguities} have been discussed in Refs.\ \cite{JS98}, 
\cite{JS99}, 
and, in more detail, in the Appendix A of \cite{DJS1}. We have recomputed, in 
an arbitrary reference frame, the various regularized versions of all the 
momentum-dependent formal ``exact divergences'' $\Delta_{31}, \Delta_{32}, 
\ldots , \Delta_{38}$ defined in Appendix A of \cite{DJS1}. These 
contributions should formally vanish, but their regularized values do not 
vanish and thereby exhibit the regularization ambiguities present at $3$~PN. 
We find (in confirmation of the result given in the Introduction section of 
Ref.~\cite{JS98}) that all the momentum-dependent regularization ambiguities 
are equivalent to adding to Eq.\ (\ref{eq17}) a term of the 
(specific\footnote{Note in particular the absence of terms mixing 
${\bf p}_1$ and ${\bf p}_2$.}) form
\be
\label{eq18}
H_{\rm 3PN}^{\rm kinetic} ({\bf x}_a , 
{\bf p}_a) = + \frac{1}{2}\, \omk \, \frac{G^3 \, m_1 \, m_2}{r_{12}^3}
\left[ {\bf p}_1^2 - 3 ({\bf n}_{12} 
\cdot {\bf p}_1)^2 + {\bf p}_2^2 - 3 
({\bf n}_{12} \cdot {\bf p}_2)^2 \right] \, , 
\ee
where $\omk$ is an arbitrary parameter. In addition to this 
``kinetic'' regularization ambiguity, it was pointed out in \cite{JS99} and 
\cite{DJS1}, that there is also a ``static'' (i.e.\ momentum-independent) 
regularization ambiguity of the form 
\begin{equation}
\label{eq19}
H_{\rm 3PN}^{\rm static} ({\bf x}_a,{\bf p}_a)
= + \omega_{\rm static} \, \frac{G^4 \, m_1^2 \, m_2^2 \, (m_1 + m_2)}{r_{12}^4} 
\, , 
\end{equation}
where $\oms$ is a second arbitrary parameter. Finally,
the 3PN (order-reduced) Hamiltonian is of the form
\be
\label{eq20}
H_{\rm 3PN}({\bf x}_a,{\bf p}_a)
= H_{\rm 3PN}^{\rm reg}
+ H_{\rm 3PN}^{\rm kinetic}
+ H_{\rm 3PN}^{\rm static} \, , 
\ee
and depends on two, up to now undetermined, real parameters
$\omk$ and $\oms$.

The problem to solve is now the following: does there exist a 
(3PN-accurate) center-of-mass vector, of the generic form,
\begin{equation}
G^i ({\bf x}_a , {\bf p}_a) = \sum_a (M_a 
({\bf x}_b , {\bf p}_b) \, x_a^i 
+ N_a 
({\bf x}_b , {\bf p}_b) \, p_a^i) \, , \label{eq21}
\end{equation}
where $M_a$ and $N_a$ are scalars that reduce to $m_a$ and 0, respectively, 
in the Newtonian approximation, such that Eqs.~(\ref{eq6})--(\ref{eq8}) are 
fulfilled (within the 3PN accuracy) when the Hamiltonian is given by 
inserting Eqs.~(\ref{eq12}), (\ref{eq14}), (\ref{eq15}) and (\ref{eq20}) in 
Eq.~(\ref{eq11})? We have tackled this problem by the method of undetermined 
coefficients, i.e. by writing the most general expressions for the successive 
PN approximations to the functions
$M_a ({\bf x}_b,{\bf p}_b)$ and $N_a ({\bf x}_b,{\bf p}_b)$,
\begin{equation}
M_a = m_a + \frac{1}{c^2}\,M_a^{\rm 1PN} 
+ \frac{1}{c^4} \, M_a^{\rm 2PN} + \frac{1}{c^6} \, M_a^{\rm 3PN} \ ;
N_a = \frac{1}{c^4} \, N_a^{\rm 2PN} 
+ \frac{1}{c^6} \, N_a^{\rm 3PN} \, , \label{eq22}
\end{equation}
as sums of scalar monomials of the form:
$c_{n_0 n_1 n_2 n_3 n_4 n_5} 
r_{12}^{-n_0} ({\bf p}_1^2)^{n_1} \, 
({\bf p}_2^2)^{n_2} \, ({\bf p}_1 \cdot 
{\bf p}_2)^{n_3} ({\bf n}_{12} \cdot 
{\bf p}_1)^{n_4} \, ({\bf n}_{12} \cdot 
{\bf p}_2)^{n_5}$,
with positive integers $n_0,\ldots,n_5$.
Besides dimensional analysis (which constrains the possible values of $n_0 , 
\ldots , n_5$ at each given PN order), and Euclidean covariance, including 
Parity symmetry, we only required time reversal symmetry (which imposes that 
$M_a$ be even, and $N_a$ odd, under ${\bf p}_a \rightarrow - 
{\bf p}_a$). We did not impose any a priori constraints on the 
mass dependence of the coefficients $c_n (m_1 , m_2)$, nor did we use the $1 
\leftrightarrow 2$ relabeling symmetry.

The 1PN approximation to $G^i$ being well-known (see, e.g., \cite{LL}),
\begin{mathletters}
\label{eq23}
\bea
M_1^{\rm 1PN}  &=&
\frac{1}{2}\frac{\pipi}{m_1}
- \frac{1}{2}\frac{Gm_1m_2}{r_{12}},
\\[2ex]
N_1^{\rm 1PN} &=& 0 \, , 
\eea
\end{mathletters}
with $M_2^{\rm 1PN}$ obtained by a $1\longleftrightarrow 2$ relabeling, we 
started looking for the most general $G^i$ at the 2PN level. At this 
level, there are 20 unknown coefficients $c_n$, and Eq.\ (\ref{eq6}) yields 40 
equations to be satisfied. We found that there is a unique 
solution\footnote{All the algebraic manipulations reported in this paper were 
done with the aid of {\textsc MATHEMATICA}.} to these redundant equations, 
namely
\begin{mathletters}
\label{eq24}
\bea
\label{eq24a}
M^{\text{2PN}}_1 &=& -\frac{1}{8}\frac{\pipip^2}{m_1^3}
+ \frac{1}{4}\frac{Gm_1m_2}{r_{12}} \left[ -5\,\frac{\pipi}{m_1^2}
- \frac{\pjpj}{m_2^2} + 7\,\frac{\pipj}{m_1m_2} + \frac{\npi\npj}{m_1m_2} 
\right]
\nonumber\\[2ex]&&
+ \frac{1}{4}\frac{Gm_1m_2}{r_{12}}\frac{G(m_1+m_2)}{r_{12}},
\\[2ex]
\label{eq24b}
N^{\text{2PN}}_1 &=& -\frac{5}{4}\, G\, \npj,
\eea
\end{mathletters}
with $M^{\text{2PN}}_2$ and $N^{\text{2PN}}_2$ obtained by a
$1\longleftrightarrow 2$ relabeling.

We have a posteriori checked that this unique ADM-gauge, 2PN 
center-of-mass vector agrees (after taking into account the shift 
${\bf x}_a^{\rm ADM}={\bf z}_a-\delta^*{\bf z}_a (z,\dot{z})$ \cite{DS88})
both with the harmonic-gauge 2PN $G^i ({\bf z}_a ,\dot{\bf z}_a)$
first derived in Ref.\ \cite{DD81b}, and with 
the Landau-Lifshitz-like \cite{LL}, ADM-gauge calculation of
$G^i({\bf x}_a , \dot{\bf x}_a)$ performed in 
Ref.\ \cite{OK89}. We have also checked that the remaining Poincar\'e-symmetry 
constraints, Eqs.\ (\ref{eq7}) and (\ref{eq8}), are also fulfilled. Concerning 
Eq.\ (\ref{eq7}), it is easy to see, in general, that it is equivalent to the 
constraint
\begin{equation}
\sum_a \, M_a ({\bf x}_b , {\bf p}_b) = 
\frac{1}{c^2} \, H ({\bf x}_b , {\bf p}_b) \, . 
\label{eq25}
\end{equation}

At the 3PN level, the most general ansatz for $M_a^{\rm 3PN}$, 
$N_a^{\rm 3PN}$, involves 78 unknown coefficients $c_n$, while 
Eq.\ (\ref{eq6}) yields 138 equations to be satisfied. The quantity 
$\omega_{\rm kinetic}$ parametrizing the momentum-dependent regularization 
ambiguity (\ref{eq18}) in the 3PN Hamiltonian enters the system of 
equations for the unknown $c_n$'s. [Indeed, it was recently noticed that 
$H_{\rm 3PN}^{\rm kinetic}$ is not separately boost-invariant 
\cite{JSWeimar}.] By contrast, the other regularization ambiguity 
(\ref{eq19}) drops out of the problem (because $H_{\rm 3PN}^{\rm 
static}$ is Galileo invariant). We found that there was a {\it unique} value 
of $\omega_{\rm kinetic}$ for which the system of equations to be satisfied 
was compatible, namely
\begin{equation}
\omega_{\rm kinetic} = \frac{41}{24} \, . \label{eq26}
\end{equation}
If $\omk\ne41/24$, the 3PN Hamiltonian does not admit a 
global Poincar\'e invariance. If $\omk=41/24$, there is a 
{\it unique} solution to Eq.\ (\ref{eq6}), namely
\begin{mathletters}
\label{eq27}
\begin{eqnarray}
\label{eq27a}
M^{\text{3PN}}_1 &=&
\frac{1}{16}\frac{\pipip^3}{m_1^5}
+ \frac{1}{16} \frac{Gm_1m_2}{r_{12}} \left[
9\,\frac{\pipip^2}{m_1^4}
+ \frac{\pjpjp^2}{m_2^4}
- 11\,\frac{\pipi\,\pjpj}{m_1^2m_2^2}
- 2\,\frac{\pipj^2}{m_1^2m_2^2}
+ 3\,\frac{\pipi\,\npj^2}{m_1^2m_2^2}
\right.\nonumber\\[2ex]&&\left.
+ 7\,\frac{\pjpj\,\npi^2}{m_1^2m_2^2}
- 12\,\frac{\pipj\,\npi\npj}{m_1^2m_2^2}
- 3\,\frac{\npi^2\npj^2}{m_1^2m_2^2} \right]
\nonumber\\[2ex]&&
+ \frac{1}{24}\frac{G^2m_1m_2}{r_{12}^2} \left[ 
(112m_1+45m_2)\frac{\pipi}{m_1^2}
+ (15m_1+2m_2)\frac{\pjpj}{m_2^2}
- \frac{1}{2}(209m_1+115m_2)\frac{\pipj}{m_1m_2}
\right.\nonumber\\[2ex]&&\left.
- (31m_1+5m_2)\frac{\npi\npj}{m_1m_2}
+ \frac{\npi^2}{m_1}
- \frac{\npj^2}{m_2}
\right]
\nonumber\\[2ex]&&
- \frac{1}{8} \frac{Gm_1m_2}{r_{12}}\frac{G^2 (m_1^2+5m_1m_2+m_2^2)}{r_{12}^2},
\\[2ex]
\label{eq27b}
N^{\text{3PN}}_1 &=& \frac{1}{8}\frac{G}{m_1m_2}
\left[ 2\,\pipj\npj - \pjpj\,\npi + 3\,\npi\npj^2 \right]
\nonumber\\[2ex]&&
+ \frac{1}{48}\frac{G^2}{r_{12}}
\left[ 19\,m_2\,\npi + \left(130\,m_1+137\,m_2\right)\npj \right].
\end{eqnarray}
\end{mathletters}
We have then checked that this unique solution does satisfy the remaining 
Poincar\'e-symmetry constraints, Eqs.\ (\ref{eq7}) and (\ref{eq8}), or, 
equivalently, Eqs.\ (\ref{eq25}) and (\ref{eq8}). It is to be noted that the 
last two momentum-dependent terms in $M_1^{\rm 3PN}$, proportional to 
$({\bf n}_{12}\cdot{\bf p}_1)^2/m_1-({\bf n}_{12}\cdot{\bf p}_2)^2/m_2$, are 
antisymmetric in the labels $1\longleftrightarrow 2$ and therefore drop out in 
the constraint (\ref{eq25}), which reads
$M_1^{\rm 3PN}+M_2^{\rm 3PN}=H_{\rm 2PN}$.
In fact, the corresponding monomials appear 
nowhere in $H_{\rm 2PN}$, but must crucially be included in $M_a^{\rm 3PN}$.

The main conclusion of this work is therefore that the necessary existence of 
a global Poincar\'e symmetry in the two-body problem {\it uniquely fixes} the 
regularization ambiguity parameter $\omega_{\rm kinetic}$ to the value 
(\ref{eq26}). The explicit realization of this Poincar\'e invariance is then 
defined by the phase-space generator $G^i ({\bf x}_a ,  
{\bf p}_a)$ defined by Eqs.\ (\ref{eq21}), (\ref{eq22}), 
(\ref{eq23}), (\ref{eq24}), and (\ref{eq27}).

Within the ADM formalism it would be very difficult to implement a 
Poincar\'e-invariant regularization procedure. (The situation is different in 
harmonic coordinates, where one can conceive a Lorentz-invariant 
regularization \cite{BF}.) It is very satisfying (and in keeping with the 
general lore about renormalization theory) that we were able to use a 
non-Poincar\'e-invariant regularization, but then, a posteriori, correct for 
it in a unique way. There remains, however, a last regularization 
ambiguity\footnote{As argued in Ref.\ \cite{JS99} this ``static'' 
regularization ambiguity seems to be linked to the breakdown of the 
possibility to use Dirac-delta-functions to model extended objects, such as 
neutron stars or black holes.}, Eq.\ (\ref{eq19}), which has all the needed 
global symmetries and cannot be fixed in this way.

\section*{Acknowledgments}

We thank L.\ Blanchet for informing us, before completion of our and his work,
that he and G.\ Faye had succeeded in determining $\omk$.  He communicated to us
the numerical value $\omk\simeq1.71$. P.J.\ gratefully acknowledges useful
discussions with Piotr Bizo\'n and Prof.\ Andrzej Staruszkiewicz.  P.J.\ and
G.S.\ thank the Institut des Hautes \'Etudes Scientifiques for hospitality
during the realization of this work.  This work was supported in part by the KBN
Grant No.\ 2 P03B 094 17 (to P.J.)  and the Max-Planck-Gesellschaft Grant No.\
02160-361-TG74 (to G.S.).

\end{document}